\documentclass[12pt,a4paper,english]{revtex4}
\usepackage[T1]{fontenc}
\usepackage[latin9]{inputenc}
\usepackage{float}
\usepackage{amsmath}
\usepackage{graphicx}
\usepackage{amssymb}
\usepackage{esint}

\makeatletter
\@ifundefined{textcolor}{}
{%
 \definecolor{BLACK}{gray}{0}
 \definecolor{WHITE}{gray}{1}
 \definecolor{RED}{rgb}{1,0,0}
 \definecolor{GREEN}{rgb}{0,1,0}
 \definecolor{BLUE}{rgb}{0,0,1}
 \definecolor{CYAN}{cmyk}{1,0,0,0}
 \definecolor{MAGENTA}{cmyk}{0,1,0,0}
 \definecolor{YELLOW}{cmyk}{0,0,1,0}
 }

\usepackage{babel}

\makeatother

\usepackage{babel}

\begin{document}

\title{Coherence revival during the attosecond electronic and nuclear quantum
photodynamics of the ozone molecule }

\author{G. J. Halász,$^{1}$ A. Perveaux,$^{2}$ B. Lasorne,$^{2}$ M. A.
Robb,$^{3}$ F. Gatti,$^{2}$ and Á. Vibók$^{4}$}

\affiliation{$^{1}$Department of Information Technology, University of Debrecen,
H-4010 Debrecen, PO Box 12, Hungary}

\affiliation{$^{2}$CTMM, Institut Charles Gerhardt Montpellier, Université Montpellier
2, F-34095 Montpellier, France}

\affiliation{$^{3}$Imperial College London, Department of Chemistry, London SW7
2AZ, UK}

\affiliation{$^{4}$Department of Theoretical Physics, University of Debrecen,
H-4010 Debrecen, PO Box 5, Hungary}

\email{vibok@phys.unideb.hu}
\begin{abstract}
A coherent superposition of two electronic states of ozone (ground
and Hartley B) is prepared with a UV pump pulse. Using the multiconfiguration
time-dependent Hartree approach, we calculate the subsequent time
evolution of the two corresponding nuclear wave packets and the coherence
between them. The resulting wave packet shows an oscillation between
the two chemical bonds. Even more interesting, the coherence between
the two electronics states reappears after the laser pulse is switched
off, which could be observed experimentally with an attosecond probe
pulse.
\end{abstract}
\maketitle

\section{Introduction}

The construction of single few-cycle ultrashort laser pulses or trains
of ultrashort pulses enables controlling different photophysical and
photochemical processes. Experimentalists can excite and probe electron
dynamics in atoms and molecules in real time \cite{Feri1,Feri2,Feri3,Smirnova1,Huis,Murnane,santra1,santra2,Kelken,elefeitosz1}.
Monitoring the subfemtosecond motion of valence electrons over a multifemtosecond
time span that results in taking real-time snapshots of ultrafast
transformations of matter. Successful theoretical and experimental
investigations of the electron dynamics of the Kr atom have been performed
recently \cite{santra1,santra2,elefeitosz1}. However, extending these
techniques to molecules remains a challenge. Problems arise because
electron dynamics in molecules often are strongly coupled to nuclear
dynamics.

For molecules, various approaches have been developed so far. In most
attophysics simulations, only the electron dynamics is treated, and
the molecular geometries (nuclear positions) are assumed to be fixed
\cite{levine1a,levine2a,levine2aa,nest1,levine3a,levine4a}. Within
this approach an arbitrarily large molecule can be examined. To achieve
this, one needs to use an ultrashort laser pulse during the probe
process. If longer probe laser pulses are applied, the nuclei have
time to move. In this situation the nuclear dynamics has to be considered
as well. For the simplest ion, H$_{2}^{+}$, or molecule, H$_{2}$,
it is easily feasible \cite{mainz1a,mainz2a,vrakking1,moshammer2,fernando1,fernando2,steffi2},
but for diatomics containing many electrons or even for polyatomics
the problem to be solved is more complex and difficult \cite{regine2,regine3,regine4}.
In the first situation (e.g. H$_{2}^{+}$ or H$_{2}$) the total time-dependent
Schrödinger equation (TDSE) can be solved numerically including explicitly
both the electronic and nuclear degrees of freedom. In contrast, the
case of many electrons or polyatomics implies to face either the problem
of electron correlation or of a large number of nuclear degrees of
freedom \cite{lenz4a}.

Recently, we proposed a nonadiabatic scheme for the description of
the coupled electron and nuclear motion in the ozone molecule \cite{Agnee}.
An initial coherent nonstationary state was prepared by two pump pulses.
It was a superposition of different weakly-bound states in the Chappuis
band \cite{schinke2} (which are populated by NIR radiation), as well
as in the Hartley band \cite{schinke2} (which is populated by the
3rd harmonic pulse). In this situation neither the electrons nor the
nuclei were in a stationary state, and we used nonadiabatic quantum
dynamics simulations. As the transition dipole moments are very different
between the ground and Hartley states compared to the ground and Chappuis
bands we had to apply significantly different intensities for the
two pump pulses not to obtain differences between the populations
of the Hartley and the Chappuis states larger than one order of magnitude.
Consequently, we used $2\times10^{11}$ and $10^{14}$ W/cm$^{2}$
intensities to populate the Hartley and Chappuis states, respectively,
which is not trivial to achieve experimentally while further probing
the system with an attosecond XUV pulse. 

However, opportunities arise to reasonably simplify the task. As we
excite only the B state of the Hartley band with a much larger intensity
pump pulse than in our previous work, the population obtained in this
state is more pronounced. The non-stationary state is a coherent superposition
of these two (ground and B) electronic states, and the motion of the
electronic wave packet can thus be probed assuming much less complicated
experimental setups than in the previous situation. 

Our original motivation was to perform a numerical simulation for
an experimentally easier situation. An interesting phenomenon emerged
from this investigation: the revival of the electronic coherence after
the pump pulse is off, which could also be probed experimentally.
The main aim of the present paper is to report this uncommon finding
that can be explained because we only coupled the X and B electronic
states, between which there is no nonadiabatic coupling and no conical
intersection. 

As in our previous work, the nuclear wave packets, the electronic
populations, the relative electronic coherence between the ground
X and B electronic states and the electron wave packet dynamics were
calculated. The time evolution of the electronic motion was plotted
in the Franck-Condon (FC) region only due to the localization of the
nuclear wave packet around this point during the first $5-6$ fs.
The electron density shows a fast oscillation pattern between both
chemical bonds, which we expect could be observed by an attosecond
probe pulse. 

The paper is organized as follows. Sec. II gives some insights into
the formalism and methods used here. Results and their discussions
are presented in Sec. III. Sec. IV is devoted to conclusions. Some
useful remarks are provided in appendix about the electronic-structure
results.

\section{Methods and Formalism}

In this section a short summary is given about the methods and formalism
used in our simulations. For more details we refer to our former paper
\cite{Agnee}.

\subsection{Time-dependent molecular Schrödinger equation}

In the adiabatic partition (beyond Born-Oppenheimer \cite{Born}),
the total molecular wave function $\Psi_{tot}(\vec{r}_{el},\vec{R},t)$
can be assumed as a sum of products of electronic wave functions,
$\psi_{el}^{k}(\vec{r}_{el};\vec{R})$, and nuclear wave packets,
$\Psi_{nuc}^{k}(\vec{R},t)$:

\begin{equation}
\Psi_{tot}(\vec{r}_{el},\vec{R},t)=\sum_{k=1}^{n}\Psi_{nuc}^{k}(\vec{R},t)\psi_{el}^{k}(\vec{r}_{el};\vec{R}).\label{eq:totwave}\end{equation}

Here $k$ denotes the $k-th$ adiabatic electronic state, $\vec{r}_{el}$
and $\vec{R}$ are the electronic and the nuclear coordinates, respectively.
We are interested in solving the coupled evolution of the nuclear
wave packets, $\Psi_{nuc}^{k}(\vec{R},t)$, by inserting the product
ansatz (\ref{eq:totwave}) into the time-dependent Schrödinger equation
of the full molecular Hamiltonian. Integrating over the electronic
coordinates one obtains the coupled nuclear Schrödinger equations:

\begin{equation}
i\hbar\frac{\partial}{\partial t}\Psi_{nuc}^{k}(\vec{R},t)=\sum_{l=1,n}H_{k,l}\Psi_{nuc}^{l}(\vec{R},t).\label{eq:schrodinger}\end{equation}
Here $H_{k,l}$ is the matrix element of the vibronic Hamiltonian,
which reads, e.g., for $n=2$,

\begin{equation}
H=\left(\begin{array}{cc}
T_{nuc}+V_{k} & K_{k,l}\\
-K_{k,l} & T_{nuc}+V_{l}\end{array}\right),\label{eq:nuc-ham}\end{equation}
where $T_{nuc}$ is the nuclear kinetic energy, $V_{k}$ ($k=1,...n$)
is the $k-th$ adiabatic potential energy and $K_{k,l}$ with $k\neq l$
is the vibronic coupling term between the $(k,l)-th$ electronic states.
The latter contains the nonadiabatic coupling term (NACT). In the
presence of an external electric field the light-matter interaction,
$-\vec{\mu}(k,l)\cdot\overrightarrow{E}(t)$ (electric dipole approximation),
where $\vec{E}(t)$ is an external field resonant between the $k-th$
and the $l-th$ states and $\vec{\mu}(k,l)$ is the $\vec{R}-$dependent
transition dipole moment, is also included in this coupling term.
In the present situation, there is no significant nonadiabatic coupling
between the ground and Hartley state, therefore $K_{k,l}$ denotes
only the light-matter interaction.

One has to solve the time-dependent nuclear Schrödinger equation given
by Eq. (\ref{eq:schrodinger}). One of the most efficient approaches
for this is the MCTDH (multiconfiguration time-dependent Hartree)
method \cite{dieter1a,dieter2a,dieter3a,dieter4a}.

The MCTDH nuclear wave packets, $\Psi_{nuc}^{k}(\vec{R},t)$, contain
all the information about the relative phases between the electronic
states. Therefore $\Psi_{nuc}^{k}(\vec{R},t)$ can also be written
as: \begin{equation}
\Psi_{nuc}^{k}(\vec{R},t)=\exp(-iW_{k}(\vec{R})t/\hbar)a_{k}(\vec{R},t).\label{eq:phase}\end{equation}
Here, $W_{k}(\vec{R})$ is the potential energy of the $k-th$ state.
The first part of this wave function is the phase factor, ($\exp(-iW_{k}(\vec{R})t/\hbar)$),
of the $k-th$ state, which oscillates very fast.

\subsection{Density Matrix}

Here we define the working formulas that are used in the next section.
Calculating them only requires the knowledge of the nuclear wave packets.

The two-dimensional nuclear density function (depending on $R_{1}$
and $R_{2}$, the two bond lengths, and integrated over $\theta$,
the bond angle) is:

\begin{center}
\begin{equation}
\left|\Psi_{nuc}^{i}(R_{1},R_{2},t)\right|^{2}=\int\Psi_{nuc}^{i}(R_{1},R_{2},\theta,t)\Psi_{nuc}^{i*}(R_{1},R_{2},\theta,t)\sin\theta d\theta.\label{eq:density}\end{equation}

\par\end{center}

The total density matrix of the molecule is defined as:

\begin{eqnarray}
\rho_{ii^{'}}(\vec{R},\vec{R}',t) & = & \left\langle \psi_{el}^{i}(\vec{r}_{el};\vec{R})\right|\left.\Psi_{tot}(\vec{r}_{el},\vec{R},t)\right\rangle \left\langle \Psi_{tot}(\vec{r}_{el},\vec{R}',t)\right|\left.\psi_{el}^{i^{'}}(\vec{r}_{el};\vec{R}')\right\rangle \label{eq:totdensity}\\
 & = & \Psi_{nuc}^{i}(\vec{R},t)\Psi_{nuc}^{i'*}(\vec{R}',t),\nonumber \end{eqnarray}
where brackets denote integration over the electronic coordinates
only.

The electronic population function of the $i-th$ state is:

\begin{equation}
P_{i}(t)=\int\rho_{ii}(\overrightarrow{R},\overrightarrow{R},t)d\overrightarrow{R}.\label{eq:population-1}\end{equation}

Analogously, the electronic relative coherence between the $i-th$
and $i^{'}-th$ electronic states can be approximated as:

\begin{equation}
C_{ii^{'}}(t)=\int\rho_{ii^{'}}(\overrightarrow{R},\overrightarrow{R},t)d\overrightarrow{R}/\sqrt{P_{i}(t)P_{i^{'}}(t)}.\label{eq:coherence1}\end{equation}

\subsection{Electronic Structure Treatment}

Here we briefly review the represention used for the electronic wave
packet. We consider only two (ground and Hartley B) electronic states.
At the FC geometry, each electronic state can be represented by its
charge density in the three-dimensional space,

\begin{eqnarray}
\rho^{i}(\vec{r},\vec{R}_{FC}) & = & N\int_{N(spin)}d\sigma_{1}d\sigma_{2}\ldots d\sigma_{N}\int_{N-1(space)}d\tau_{2}\ldots d\tau_{N}\label{eq:charge-density}\\
 &  & \left|\psi_{el}^{i}(\vec{r}_{1}=\vec{r},\sigma_{1},\vec{r}_{2},\sigma_{2},...,\vec{r}_{N},\sigma_{N};\vec{R}_{FC})\right|^{2}.\nonumber \end{eqnarray}
Here $i=X$ or $B$. It is often called the one-electron density,
although rigorously, it is N times the one-electron density summed
over both spin states of electron 1. It is defined as the density
of probability of finding one among N electrons in any spin state
(up or down) at point $\vec{r}\equiv(x,y,z)$ and time $t$ for the
molecule in state X and B, respectively, and geometry $\vec{R}_{FC}$.

The transition density between states X and B is defined in the three-dimensional
space as:

\begin{eqnarray}
\gamma^{XB}(\vec{r};\vec{R}_{FC}) & = & N\int_{N(spin)}d\sigma_{1}d\sigma_{2}\ldots d\sigma_{N}\int_{N-1(space)}d\tau_{2}\ldots d\tau_{N}\label{eq:transition-density}\\
 & \times & \psi_{el}^{X*}(\vec{r}_{1}=\vec{r},\sigma_{1},\vec{r}_{2},\sigma_{2},...,\vec{r}_{N},\sigma_{N};\vec{R}_{FC})\nonumber \\
 & \times & \psi_{el}^{B}(\vec{r}_{1}=\vec{r},\sigma_{1},\vec{r}_{2},\sigma_{2},...,\vec{r}_{N},\sigma_{N};\vec{R}_{FC}).\nonumber \end{eqnarray}
It is a measure of the interference between both states. The total
molecular wave packet observed at a fixed geometry, here at the FC
point, is a coherent mixture of both electronic states, whereby the
time-dependent coefficients are the nuclear wave packets at the FC
point:

\begin{eqnarray}
 &  & \Psi_{mol}(\vec{r}_{1},\sigma_{1},\vec{r}_{2},\sigma_{2},...,\vec{r}_{N},\sigma_{N};\vec{R}_{FC},t)=\label{eq:tot-wave}\\
 &  & \Psi_{nuc}^{X}(\vec{R}_{FC},t)\psi_{el}^{X}(\vec{r}_{1},\sigma_{1},\vec{r}_{2},\sigma_{2},...,\vec{r}_{N},\sigma_{N};\vec{R}_{FC})\nonumber \\
 & + & \Psi_{nuc}^{B}(\vec{R}_{FC},t)\psi_{el}^{B}(\vec{r}_{1},\sigma_{1},\vec{r}_{2},\sigma_{2},...,\vec{r}_{N},\sigma_{N};\vec{R}_{FC})\nonumber \end{eqnarray}

Thus, the corresponding total time-dependent charge density reads:

\begin{eqnarray}
 &  & \rho^{tot}(\vec{r},t;\vec{R}_{FC})=|\Psi_{nuc}^{X}(\vec{R}_{FC},t)|^{2}\rho^{X}(\vec{r};\vec{R}_{FC})+|\Psi_{nuc}^{B}(\vec{R}_{FC},t)|^{2}\rho^{B}(\vec{r};\vec{R}_{FC})\nonumber \\
 & + & 2Re\Psi_{nuc}^{X*}(\vec{R}_{FC},t)\Psi_{nuc}^{B}(\vec{R}_{FC},t)\gamma^{XB}(\vec{r};\vec{R}_{FC}).\label{eq:tot-density}\end{eqnarray}

Now, we define the excited-state differential charge density at the
FC point as the difference of the total charge density between the
excited state B and the ground state %
\footnote{Some additional explanations on the definition chosen for the differential
charge density are provided in Appendix.%
}:

\begin{eqnarray}
 &  & \Delta\rho^{B}(\vec{r},t;\vec{R}_{FC})=\rho^{tot}(\vec{r},t;\vec{R}_{FC})-[|\Psi_{nuc}^{X}(\vec{R}_{FC},t)|^{2}+|\Psi_{nuc}^{B}(\vec{R}_{FC},t)|^{2}]\rho^{X}(\vec{r};\vec{R}_{FC})\nonumber \\
 & = & |\Psi_{nuc}^{B}(\vec{R}_{FC},t)|^{2}[\rho^{B}(\vec{r};\vec{R}_{FC})-\rho^{X}(\vec{r};\vec{R}_{FC})]+2Re\Psi_{nuc}^{X*}(\vec{R}_{FC},t)\Psi_{nuc}^{B}(\vec{R}_{FC},t)\gamma^{XB}(\vec{r};\vec{R}_{FC})\nonumber \\
 & = & |\Psi_{nuc}^{B}(\vec{R}_{FC},t)|^{2}\triangle\rho^{B}(\vec{r};\vec{R}_{FC})+2Re\Psi_{nuc}^{X*}(\vec{R}_{FC},t)\Psi_{nuc}^{B}(\vec{R}_{FC},t)\gamma^{XB}(\vec{r};\vec{R}_{FC}),\label{eq:ex-state-diff-cd}\end{eqnarray}
where $\triangle\rho^{B}(\vec{r};\vec{R}_{FC})=\rho^{B}(\vec{r};\vec{R}_{FC})-\rho^{X}(\vec{r};\vec{R}_{FC})$.

\section{Results and Discussion}

In our present work only two electronic states of ozone are involved
in the numerical simulations. The gound state X with $^{1}A_{1}$
symmetry and the highly-excited B state in the Hartley band with $^{1}B_{2}$
symmetry. In Fig. 1 we show a one-dimensional cut along the O - O
bond through the potential energy surfaces (PESs) of both electronic
states. We note here, as there is no nonadiabatic coupling between
these two states, that the adiabatic and diabatic energies are identical.
A UV linearly-polarized Gaussian laser pump pulse was used to prepare
a coherent superposition of the two stationary - the ground X and
the populated B - electronic states. The center wavelength and the
intensity of the pulse are $260$ nm and $10^{13}$ W/cm$^{2}$, respectively.
The FWHM is $3$ fs. The PESs and $\vec{R}$-dependent dipole moments
occurring in the radiative coupling terms were taken from Refs. \cite{schinke1,schinke2,schinke3}.

The FC point has $C_{2v}$ symmetry. As a consequence, only the \emph{y}-component
($B_{2}$) of the transition dipole between the ground state X ($^{1}A_{1}$)
and Hartley B ($^{1}B_{2}$) is nonzero. Therefore the only effective
polarization of the electric field is $y$ (see upper panel on Fig.2).

\begin{center}
\begin{figure}
\begin{centering}
\includegraphics[width=8.5cm]{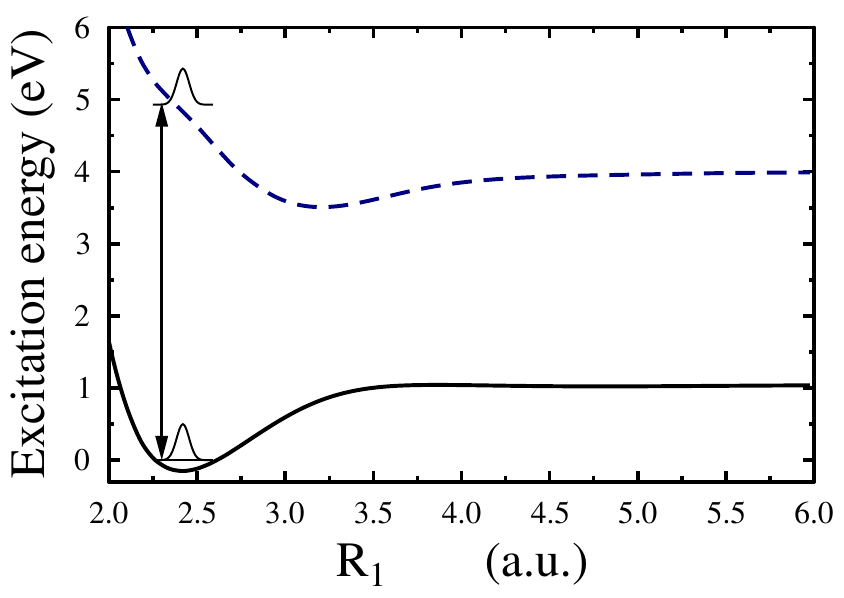} 
\par\end{centering}

\caption{(Color online) The potential energy surfaces of ozone as functions
of the dissociation coordinate: ground state (X, solid line) and Hartley
state (B, dashed line), the arrow denotes excitation of the B state. }

\end{figure}

\par\end{center}

\begin{figure}
\begin{centering}
\includegraphics[width=8.5cm]{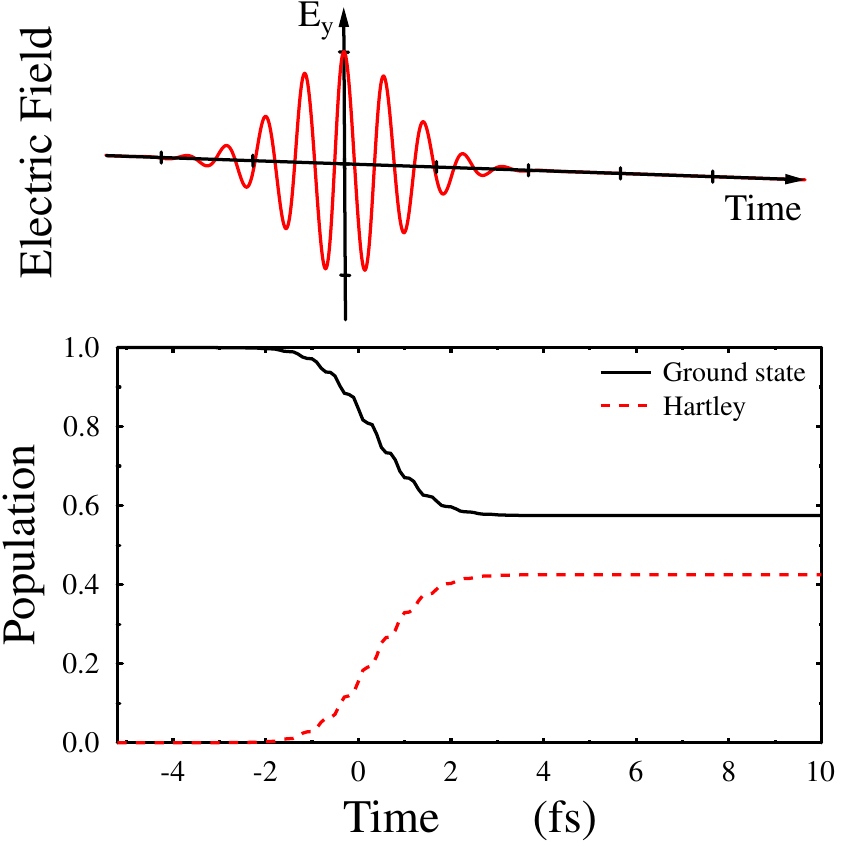} 
\par\end{centering}

\caption{(Color online) Upper panel: The applied electric field. Lower panel:
Time evolution of the diabatic populations on the ground ($X$) and
diabatic excited ($B$) states. }

\end{figure}

\begin{figure}
\begin{centering}
\includegraphics[width=8.5cm]{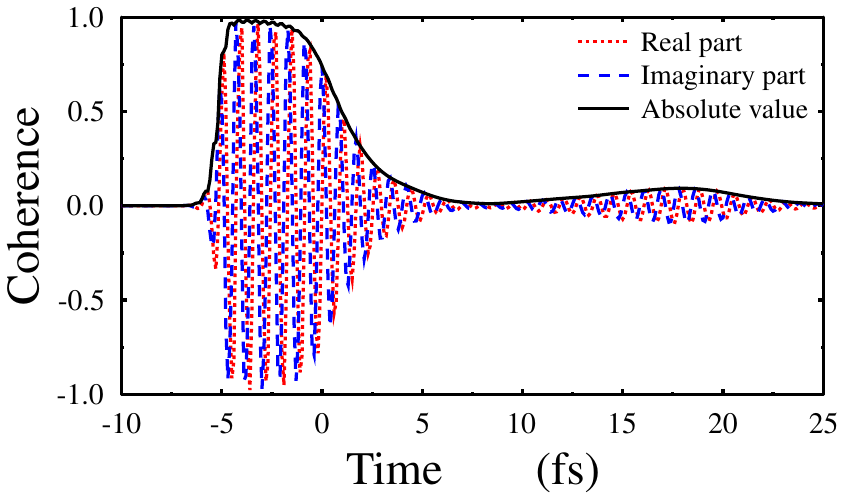} 
\par\end{centering}

\caption{(Color online) Relative electronic coherence as a function of time.
The real, the imaginary parts and the absolute value of the relative
electronic coherence between the ground ($X$) and Hartley ($B$)
states. }

\end{figure}

\begin{figure}[H]
\noindent \raggedright{}(a)\includegraphics[width=0.2\textwidth]{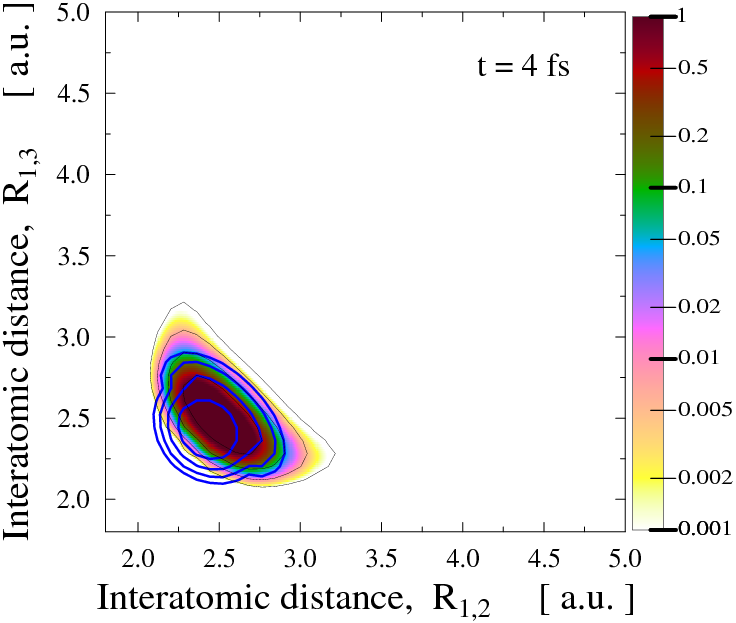}\hfill{}(b)\includegraphics[width=0.2\textwidth]{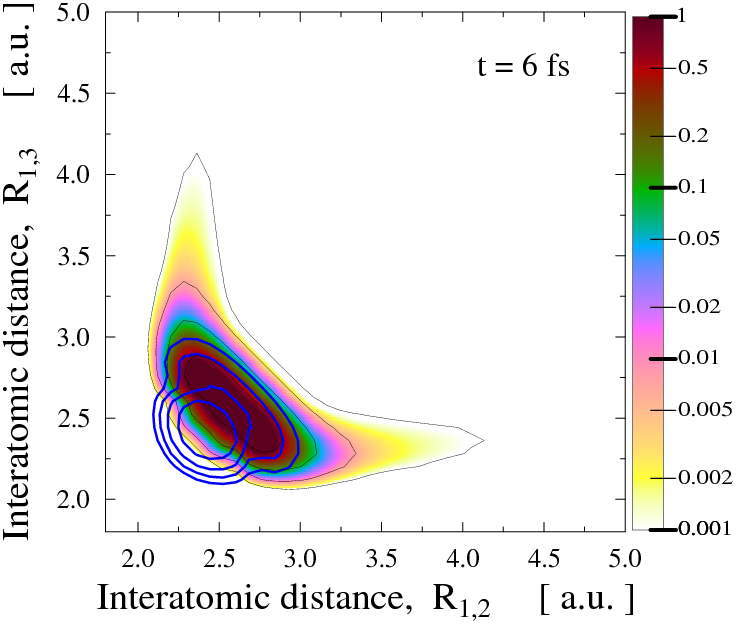}\hfill{}(c)\includegraphics[width=0.2\textwidth]{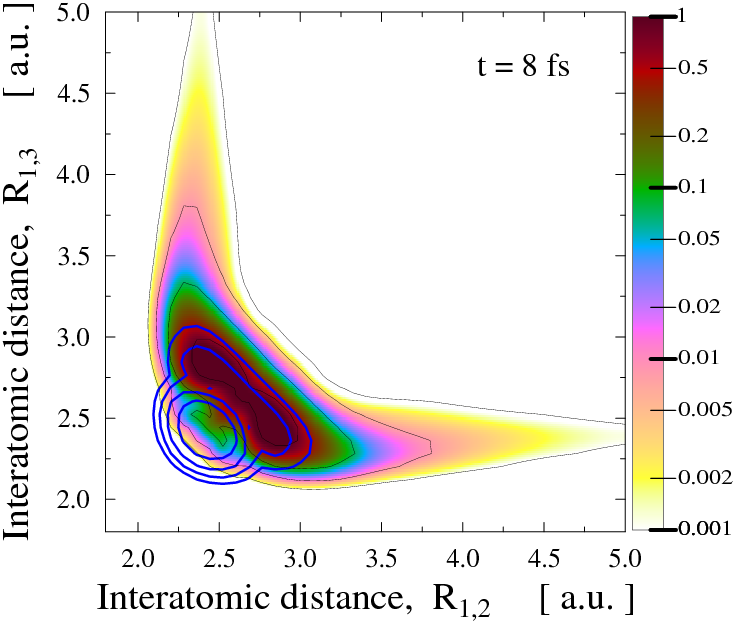}\hfill{}(d)\includegraphics[width=0.2\textwidth]{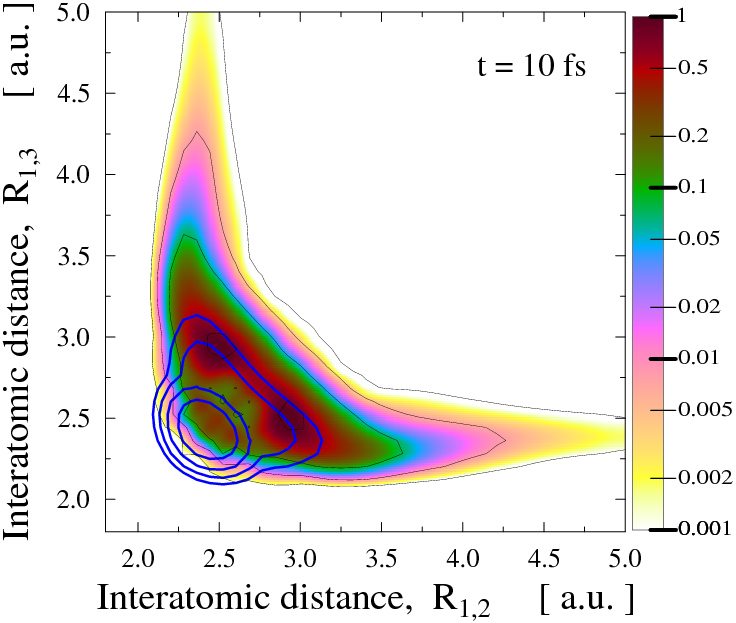}\hfill{}(e)\includegraphics[width=0.2\textwidth]{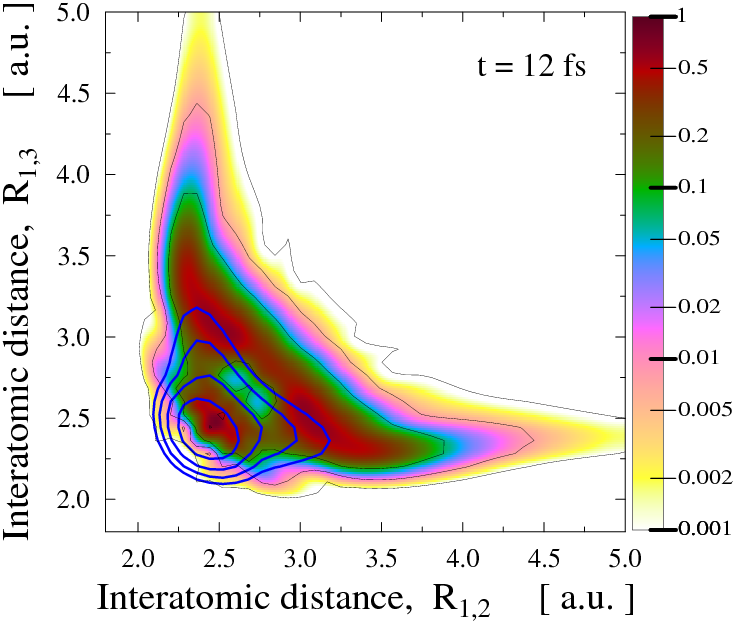}\hfill{}(f)\includegraphics[width=0.2\textwidth]{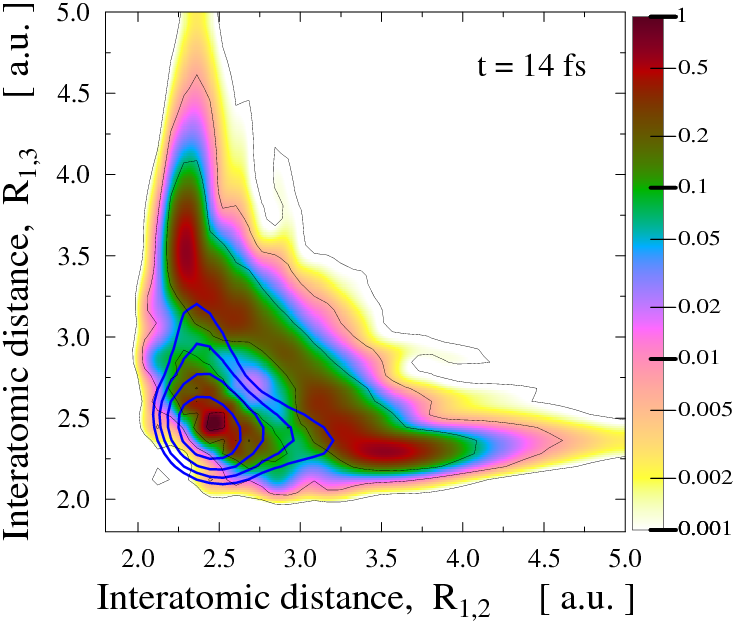}\hfill{}(g)\includegraphics[width=0.2\textwidth]{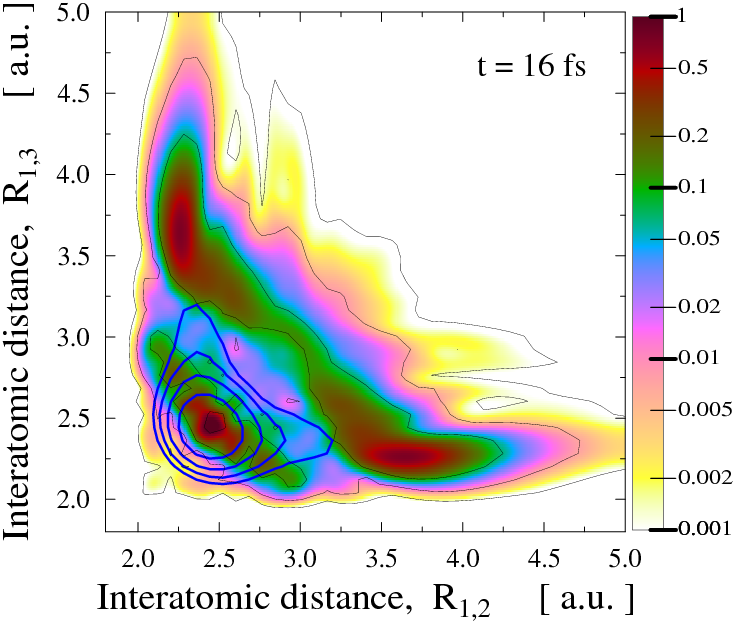}\hfill{}(h)\includegraphics[width=0.2\textwidth]{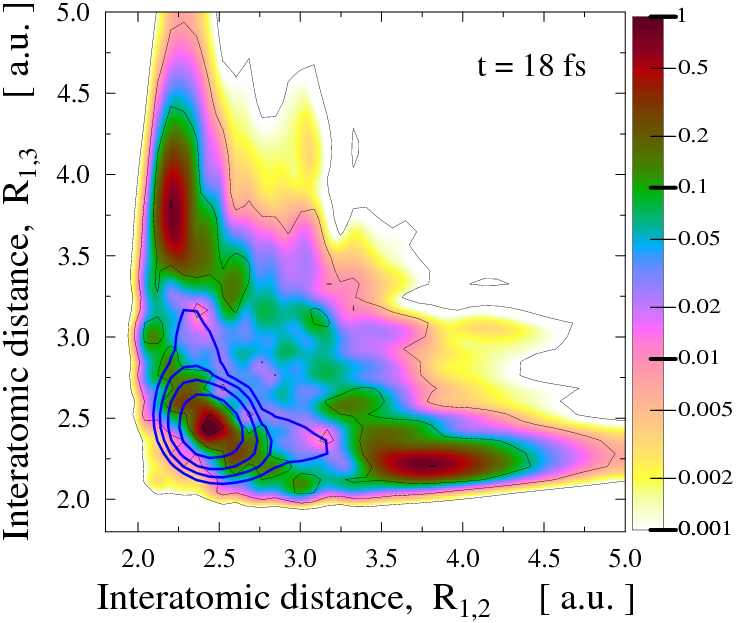}\hfill{}(i)\includegraphics[width=0.2\textwidth]{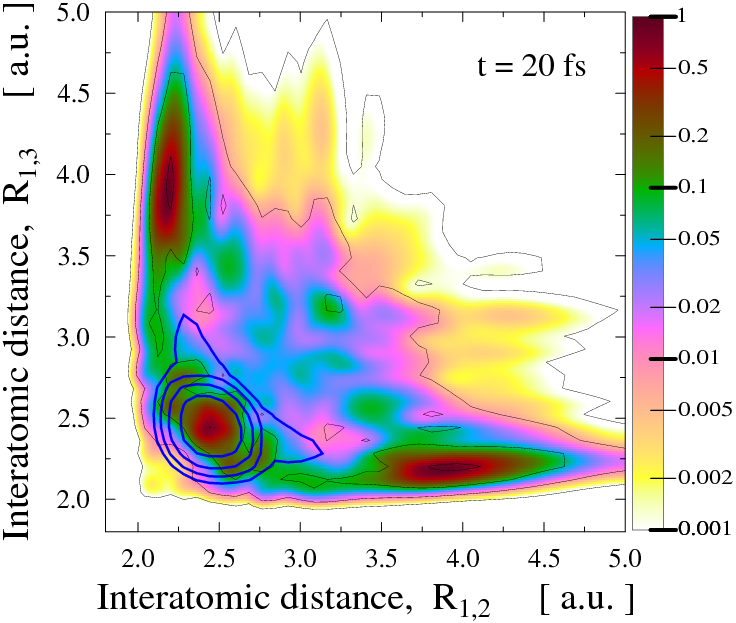}\hfill{}(j)\includegraphics[width=0.2\textwidth]{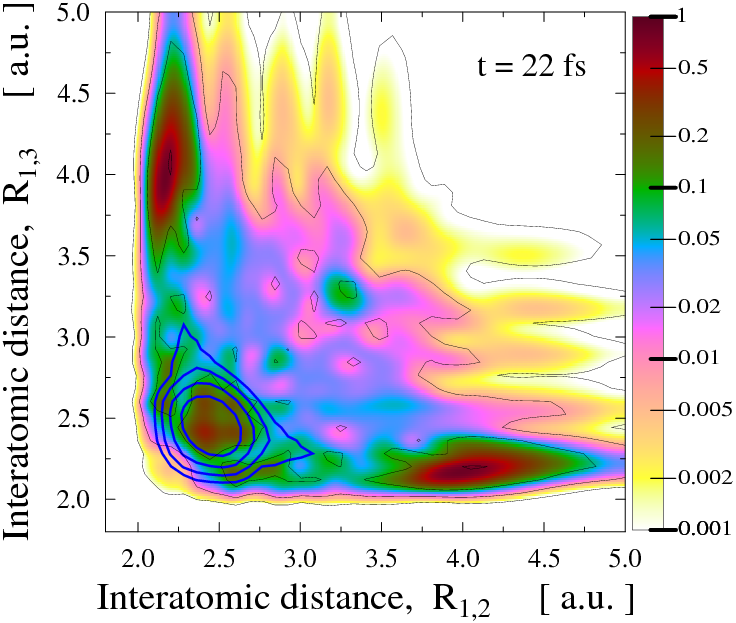}\hfill{}(k)\includegraphics[width=0.2\textwidth]{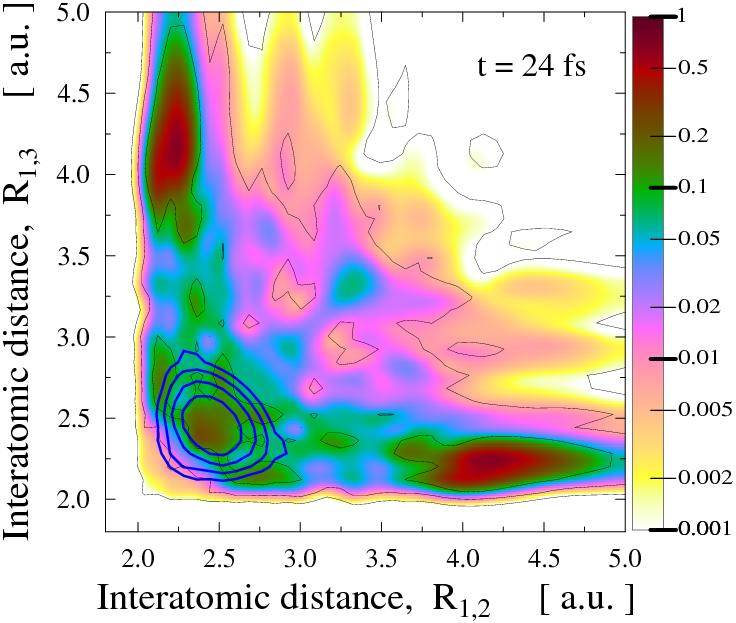}\hfill{}(l)\includegraphics[width=0.2\textwidth]{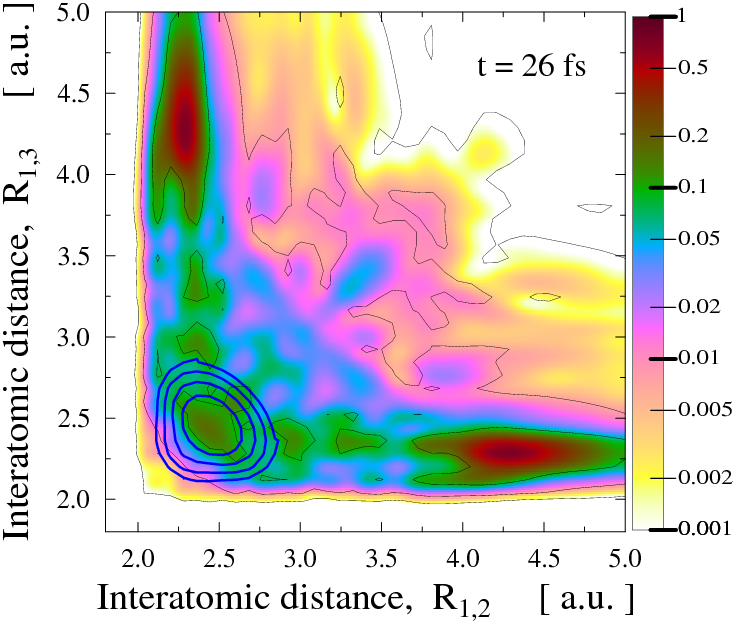}{\footnotesize \caption{{\footnotesize Snapshots of the time evolution of the nuclear wavepacket
density along both O - O bonds. }}
}{\footnotesize \par}

\end{figure}

\begin{figure}
\begin{centering}
\includegraphics[width=8.5cm]{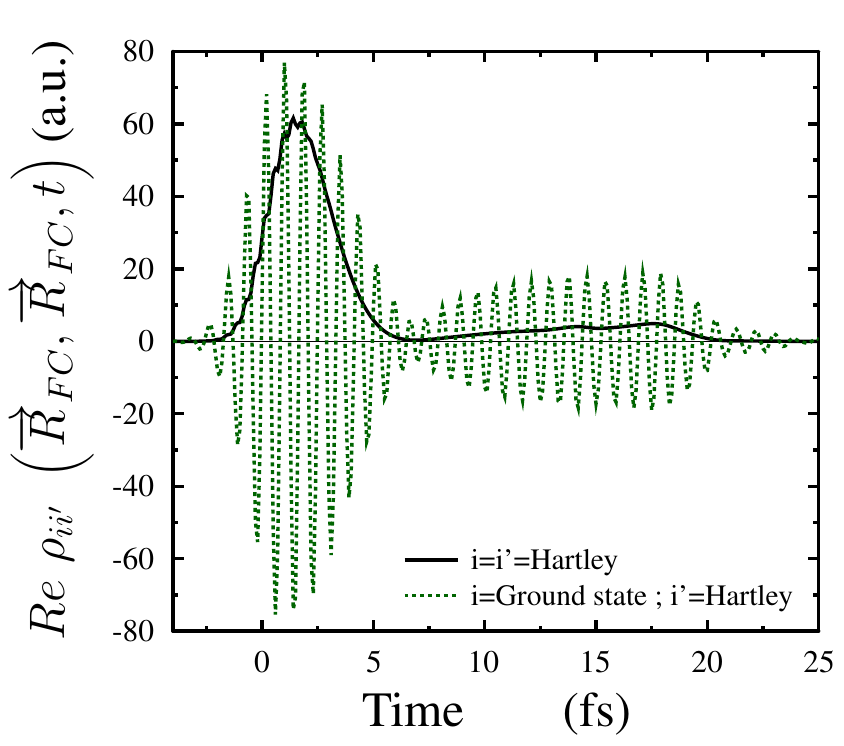} 
\par\end{centering}

\caption{(Color online) Local population density for state B (black) and real
part of the interference (last) term in Eq. (\ref{eq:ex-state-diff-cd})
(dashed green) at the FC point as functions of time.}

\end{figure}

\begin{figure}
\begin{centering}
\includegraphics[width=1\textwidth]{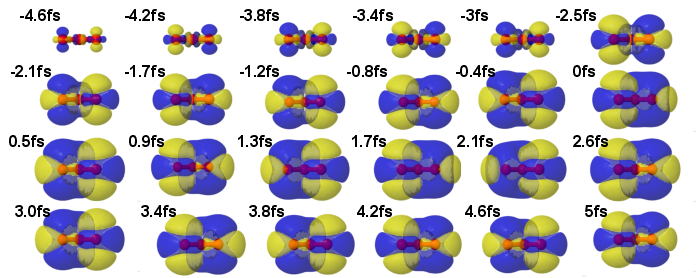} 
\par\end{centering}

\caption{(Color online) Time evolution of the excited differential electronic
charge density, Eq. (\ref{eq:ex-state-diff-cd}), at the FC geometry
(side view). Dark (blue): hole; light (yellow): electron. }

\end{figure}

In the lower panel of Fig. 2 the total populations against time, see
Eq. (\ref{eq:population-1}), are displayed in the ground and Hartley
B states up to t=$10$ fs (note that they stay constant up to the
end of the simulation, at t=$35$ fs). The Hartley B state absorbs
very strongly due to the large value of the transition dipole moment
with the ground state \cite{gabriel}. Between the ($-2$, $2$) fs
interval the population grows continuously, then reaches its maximum
and remains at this value throughout the studied time period. The
B state is populated with a yield of about 40\%. The laser intensity
($10^{13}$ W/cm$^{2}$) is thus large enough to transfer near half
of the ammount of the wave packet from the ground state to the B state.

Fig. 3 shows the electronic relative coherence, Eq. (\ref{eq:coherence1}),
between the ground and B states. In the first time period the coherence
increases very fast and reaches its maximum. It retains this value
for $3$ - $4$ fs, which is approximately equivalent to the duration
of the laser pulse and then it decays during the next $6$ - $7$
fs. However, this is not the end of the process: a few femtoseconds
later ($\sim5$ fs), the coherence reappears in contrast with what
was observed in Ref. \cite{Agnee}. This revival of coherence proves
that we have created, to some extent, a \textquotedbl{}true\textquotedbl{}
coherent superposition in that it is not forced by the presence of
an external field. This phenomenon could certanly be enhanced experimentally
by optimizing the parameters of the laser pulse. 

This revival of electronic coherence is interesting because the pump
pulse is already off. This implies that the wave packet oscillates
in the B state and then goes back to the FC region where it is still
coherent with the part left in the ground state. To understand this
more deeply we have analysed the nuclear density function, Eq. (\ref{eq:density}).
Results are illustrated in Fig. 4 with snapshots from the structure
of the nuclear wave packet density $\left|\Psi_{nuc}^{i}(R_{1},R_{2},t)\right|^{2}$
at different times. It is seen that a part of the nuclear wave packet
stays trapped on the symmetric ridge of the B potential energy surface,
where both O - O bonds increase synchronously. A valley-ridge inflection
point occurs, where the nuclear wave packet splits into three components.
One part is bound to come back to the FC region, while the rest dissociates
along either of both equivalent channels.

The local population of the Hartley B state at the FC point (see Fig.
5) has also been computed. We are again in the same situation as in
Ref. \cite{Agnee}, namely, state B is populated significantly only
during the first $\sim5$ fs time interval over which the molecule
remains around the FC region (at least approximately). However, in
this case one part of the nuclear wave packet returns back here again
later on.

The total differential charge density at the FC point, Eq. (\ref{eq:ex-state-diff-cd}),
was obtained from electronic wave functions calculated at the SA-3-CAS(18,12)/STO-3G
level of theory using a development version of the Gaussian program
\cite{gaussian}. We observed no qualitative difference of these when
increasing the basis set to aug-cc-pVQZ or when adding dynamic electron
correlation at the MRCI level of theory using the Molpro program \cite{molpro-1}.

We limited again our discussion of the electron dynamics to the FC
region only due to the localization of the nuclear wave packet around
this point during the first $5-6$ fs. We see on Fig. 6 the oscillation
of the electronic charge density from one bond to another with a period
of 0.8 fs. The resulting electronic wave packet is thus a coherent
superposition of two chemical structures, O$\cdots$ O$_{2}$ and
O$_{2}$$\cdots$ O, each having an excess or lack of electron densities
on one or the other bond. The subfemtosecond oscillation between both
structures at the FC geometry prefigures that the dissociation of
ozone could be controlled by modulating the electron density on the
attosecond time scale.

\section{Conclusions}

In summary, we have performed numerical simulations of the coupled
electron and nuclear motion in the ozone molecule on the attosecond
time scale. An initial coherent nonstationary state was created as
a coherent superposition of the ground and excited Hartley B states.
The MCTDH approach was applied to solve the dynamical Schrödinger
equation for the nuclei in the framework of the time-dependent adiabatic
partition including the light-matter interaction (electric dipole
approximation).

A reasonably large electronic coherence has been obtained between
the ground and Hartley B states during a short $5$ fs time interval.
However after this time an interesting phenomenon emerges. After the
coherence decays within a certain period of time, a few femtosecond
later, it appears again. Nuclear wave packet calculations support
that we are presently in a situation where bifurcating reaction paths
and valley-ridge inflection points are explored on the excited-state
potential energy surface. The electronic motion during the first $5-6$
fs shows an oscillation of the electronic charge density from one
bond to another with a period of $0.8$ fs. It is to be expected that
this motion can be probed experimentally by an attosecond XUV pulse.

\section*{Acknowledgements}

The authors would like to thank F. Krausz, R. Kienberger and M. Jobst
for support and for fruitful discussions. We acknowledge R. Schinke
for providing the potential energy surfaces and the transition dipole
moment and H.-D. Meyer for fruitful discussions. The authors also
acknowledge the TÁMOP 4.2.2.C-11/1/KONV-2012-0001 project. Á.V. acknowledges
the OTKA (NN103251). Financial support by the CNRS-MTA is greatfully
acknowledged.

\section{Appendix}

Starting from Eq. (\ref{eq:totdensity}) and performing further integration
over the coordinates of the {}``last'' electron and over the coordinates
of the nuclei leads to

\begin{eqnarray*}
\underbrace{\int_{(\vec{r})}\int_{(\vec{R})}\rho^{tot}(\vec{r},t,\vec{R})d\tau dV}_{=1} & = & \underbrace{\int_{(\vec{R})}|\Psi_{nuc}^{X}(\vec{R},t)|^{2}dV}_{P^{X}(t)}\underbrace{\int_{(\vec{r})}\rho^{X}(\vec{r};\vec{R})d\tau}_{=1}\\
 & + & \underbrace{\int_{(\vec{R})}|\Psi_{nuc}^{B}(\vec{R},t)|^{2}dV}_{P^{B}(t)}\underbrace{\int_{(\vec{r})}\rho^{B}(\vec{r};\vec{R})d\tau}_{=1}\\
 & + & 2Re\underbrace{\int_{(\vec{R})}\Psi_{nuc}^{X*}(\vec{R},t)\Psi_{nuc}^{B}(\vec{R},t)dV}_{S^{XB}(t)}\underbrace{\int_{(\vec{r})}\gamma^{XB}(\vec{r};\vec{R})d\tau}_{=0},\end{eqnarray*}
where $P^{X}(t)$ and $P^{B}(t)$ are the populations of states X
and B, respectively, at time t. $S^{XB}(t)$, the overlap of the nuclear
wave packets on states X and B, is a measure of the global coherence
between states X and B for all geometries. This shows that the interference
term (involving the coherence and the transition density) does not
directly contribute to the probability of finding the molecule in
a given state (it does indirectly though, by having an effect on the
time evolution of the populations).

Now, let us turn to Eq. (\ref{eq:ex-state-diff-cd}). Assuming that
the effect of the coupling with the laser pump pulse affects only
the electrons for the duration of the observation, then there is no
transfer of local population density from $\vec{R}_{FC}$ to other
values of $\vec{R}$. As long as this approximation holds, then $|\Psi_{nuc}^{X}(\vec{R}_{FC},t<0)|^{2}=|\Psi_{nuc}^{B}(\vec{R}_{FC},t>0)|^{2}+|\Psi_{nuc}^{X}(\vec{R}_{FC},t>0)|^{2}$
(where the pulse is switched on at t=0) and $\Delta\rho^{B}(\vec{r},t>0;\vec{R}_{FC})=\rho^{tot}(\vec{r},t>0;\vec{R}_{FC})-\rho^{tot}(\vec{r},t<0;\vec{R}_{FC})$,
which thus is a measure of the change of charge density due to the
pulse.

We note here: $(i)$ At the FC point the symmetry point group is $C_{2v}$.
By construction, charge densities are $A_{1}$(totally symmetric).
However, because the X and B states have $A_{1}$ and $B_{2}$ symmetries,
the transition density is $B_{2}$ (antisymmetric with respect to
the $C_{2}$ axis and the left-right mirror plane); $(ii)$ $\Delta\rho^{B}(\vec{r},\vec{R}_{FC})>0$
means a gain of electron density, whereas $\Delta\rho^{B}(\vec{r},\vec{R}_{FC})<0$
means a loss of electron density, i.e., a gain of hole density; $(iii)$
The sign of $\gamma^{XB}(\vec{r};\vec{R_{FC}})$ can be positive (constructive
interference) or negative (destructive interference). In practice,
it is not well-defined because the signs of the electronic states
are arbitrary (in fact their phases but they are chosen real-valued).
However, this does not matter in practice, because this term has $B_{2}$
symmetry, and both terminal oxygen atoms are equivalent through permutation.

\end{document}